\documentclass[aps,prl,twocolumn,superscriptaddress]{revtex4-2}
\usepackage{eurosym}
\usepackage{amsmath}
\usepackage{amssymb}
\usepackage{graphicx}

\begin{document}

\title{Taming of free volume in statistical mechanics of the hard disks model%
}
 \author{Victor M. Pergamenshchik}
\email[Contact author: ]{victorpergam@cft.edu.pl}
\affiliation{Institute of Physics, National Academy of Sciences of Ukraine,
prospekt Nauky, 46, Kyiv 03039, Ukraine\\ Center for Theoretical Physics, Polish Academy of Sciences, Al. Lotników 32/46, 02-668, Warsaw, Poland} 
\author{Taras Bryk}
\affiliation{Yukhnovskii Institute for Condensed Matter Physics, National Academy of Sciences
of Ukraine, 1 Svientsitskii Str., Lviv 79011, Ukraine}
\affiliation{Institute of Applied Mathematics and Fundamental Sciences, Lviv National Polytechnic University, 79013 Lviv, Ukraine}
\author{Andrij Trokhymchuk}
\affiliation{Yukhnovskii Institute for Condensed Matter Physics, National Academy of Sciences
of Ukraine, 1 Svientsitskii Str., Lviv 79011, Ukraine} 
\affiliation{Faculty of Chemistry and Chemical Technology, University of Ljubljana, Ve\v{c}na pot 113, 1000 Ljubljana, Slovenia}
\date{\today }

\begin{abstract}
We turn the long time puzzle of the free volume, known for its highly
irregular form, into exact analytical formulae and develop statistical
mechanics of the hard disk model. The free volume is exactly expressed in
terms of the intersection areas of up to five exclusion circles, which can
be computed analytically as functions of disk coordinates. In turn, the free
volume determines the partition function and entropy. The partition function
is shown to factorize into a product of free volumes and admits two exact
limiting forms corresponding to gaslike and liquidlike regimes. From this
construction, using Monte Carlo-generated disk coordinates, the entropy and
pressure are obtained analytically and recover the known equation of state
of hard disks in almost entire density range up to the close packing. At
intermediate densities, the theory reveals a mixed liquid regime associated
with defect formation preceding the hexagonal ordering. The intersection
area of five disks emerges as a scalar measure of the local hexagonal order.
The theory can be directly adopted for the hard sphere model.
\end{abstract}

\maketitle

\affiliation{Institute of Physics, National Academy of Sciences of Ukraine,
prospekt Nauky, 46, Kyiv 03039, Ukraine\\ Center for Theoretical Physics, Polish Academy of Sciences, Al. Lotników 32/46, 02-668, Warsaw, Poland}

\affiliation{Yukhnovskii Institute for Condensed Matter Physics, National Academy of Sciences
of Ukraine, 1 Svientsitskii Str., Lviv 79011, Ukraine} %
\affiliation{Institute of Applied Mathematics and Fundamental Sciences, Lviv
National Polytechnic University, 79013 Lviv, Ukraine}

\affiliation{Yukhnovskii Institute for Condensed Matter Physics, National Academy of Sciences
of Ukraine, 1 Svientsitskii Str., Lviv 79011, Ukraine} \affiliation{Faculty
of Chemistry and Chemical Technology, University of Ljubljana, Ve\v{c}na pot
113, 1000 Ljubljana, Slovenia}

\affiliation{Institute of Physics, National Academy of Sciences of Ukraine,
prospekt Nauky, 46, Kyiv 03039, Ukraine\\ Center for Theoretical Physics, Polish Academy of Sciences, Al. Lotników 32/46, 02-668, Warsaw, Poland}

\affiliation{Institute for Condensed Matter Physics, National Academy of Sciences
of Ukraine, 1 Svientsitskii Str., Lviv 79011, Ukraine} 
\affiliation{Institute of Applied Mathematics and Fundamental Sciences, Lviv
National Polytechnic University, 79013 Lviv, Ukraine}

\affiliation{Institute for Condensed Matter Physics, National Academy of Sciences
of Ukraine, 1 Svientsitskii Str., Lviv 79011, Ukraine} 
\affiliation{Faculty
of Chemistry and Chemical Technology, University of Ljubljana, Ve\v{c}na pot
113, 1000 Ljubljana, Slovenia}

\affiliation{Institute of Physics, National Academy of Sciences of Ukraine,
prospekt Nauky, 46, Kyiv 03039, Ukraine\\ Center for Theoretical Physics, Polish Academy of Sciences, Al. Lotników 32/46, 02-668, Warsaw, Poland}

\affiliation{Institute for Condensed Matter Physics, National Academy of Sciences
of Ukraine, 1 Svientsitskii Str., Lviv 79011, Ukraine} 
\affiliation{Institute of Applied Mathematics and Fundamental Sciences, Lviv
National Polytechnic University, 79013 Lviv, Ukraine}

\affiliation{Institute for Condensed Matter Physics, National Academy of Sciences
of Ukraine, 1 Svientsitskii Str., Lviv 79011, Ukraine} 
\affiliation{Faculty
of Chemistry and Chemical Technology, University of Ljubljana, Ve\v{c}na pot
113, 1000 Ljubljana, Slovenia}

\affiliation{Institute of Physics, National Academy of Sciences of Ukraine,
prospekt Nauky, 46, Kyiv 03039, Ukraine\\ Center for Theoretical Physics, Polish Academy of Sciences, Al. Lotników 32/46, 02-668, Warsaw, Poland}

\affiliation{Institute for Condensed Matter Physics, National Academy of Sciences
of Ukraine, 1 Svientsitskii Str., Lviv 79011, Ukraine} 
\affiliation{Institute of Applied Mathematics and Fundamental Sciences, Lviv
National Polytechnic University, 79013 Lviv, Ukraine}

\affiliation{Institute for Condensed Matter Physics, National Academy of Sciences
of Ukraine, 1 Svientsitskii Str., Lviv 79011, Ukraine} 
\affiliation{Faculty
of Chemistry and Chemical Technology, University of Ljubljana, Ve\v{c}na pot
113, 1000 Ljubljana, Slovenia}

\affiliation{Institute of Physics, National Academy of Sciences of Ukraine,
prospekt Nauky, 46, Kyiv 03039, Ukraine\\ Center for Theoretical Physics, Polish Academy of Sciences, Al. Lotników 32/46, 02-668, Warsaw, Poland}

\affiliation{Institute for Condensed Matter Physics, National Academy of Sciences
of Ukraine, 1 Svientsitskii Str., Lviv 79011, Ukraine} 
\affiliation{Institute of Applied Mathematics and Fundamental Sciences, Lviv
National Polytechnic University, 79013 Lviv, Ukraine}

\affiliation{Institute for Condensed Matter Physics, National Academy of Sciences
of Ukraine, 1 Svientsitskii Str., Lviv 79011, Ukraine} 
\affiliation{Faculty
of Chemistry and Chemical Technology, University of Ljubljana, Ve\v{c}na pot
113, 1000 Ljubljana, Slovenia}

\affiliation{Institute of Physics, National Academy of Sciences of Ukraine,
prospekt Nauky, 46, Kyiv 03039, Ukraine\\ Center for Theoretical Physics, Polish Academy of Sciences, Al. Lotników 32/46, 02-668, Warsaw, Poland}

\affiliation{Institute for Condensed Matter Physics, National Academy of Sciences
of Ukraine, 1 Svientsitskii Str., Lviv 79011, Ukraine} 
\affiliation{Institute of Applied Mathematics and Fundamental Sciences, Lviv
National Polytechnic University, 79013 Lviv, Ukraine}

\affiliation{Institute for Condensed Matter Physics, National Academy of Sciences
of Ukraine, 1 Svientsitskii Str., Lviv 79011, Ukraine} 
\affiliation{Faculty
of Chemistry and Chemical Technology, University of Ljubljana, Ve\v{c}na pot
113, 1000 Ljubljana, Slovenia}

\affiliation{Institute of Physics, National Academy of Sciences of Ukraine,
prospekt Nauky, 46, Kyiv 03039, Ukraine\\ Center for Theoretical Physics, Polish Academy of Sciences, Al. Lotników 32/46, 02-668, Warsaw, Poland}

\affiliation{Institute for Condensed Matter Physics, National Academy of Sciences
of Ukraine, 1 Svientsitskii Str., Lviv 79011, Ukraine} 
\affiliation{Institute of Applied Mathematics and Fundamental Sciences, Lviv
National Polytechnic University, 79013 Lviv, Ukraine}

\affiliation{Institute for Condensed Matter Physics, National Academy of Sciences
of Ukraine, 1 Svientsitskii Str., Lviv 79011, Ukraine} 
\affiliation{Faculty
of Chemistry and Chemical Technology, University of Ljubljana, Ve\v{c}na pot
113, 1000 Ljubljana, Slovenia}

\affiliation{Institute of Physics, National Academy of Sciences of Ukraine,
prospekt Nauky, 46, Kyiv 03039, Ukraine\\ Center for Theoretical Physics, Polish Academy of Sciences, Al. Lotników 32/46, 02-668, Warsaw, Poland}

\affiliation{Institute for Condensed Matter Physics, National Academy of Sciences
of Ukraine, 1 Svientsitskii Str., Lviv 79011, Ukraine} 
\affiliation{Institute of Applied Mathematics and Fundamental Sciences, Lviv
National Polytechnic University, 79013 Lviv, Ukraine}

\affiliation{Institute for Condensed Matter Physics, National Academy of Sciences
of Ukraine, 1 Svientsitskii Str., Lviv 79011, Ukraine} 
\affiliation{Faculty
of Chemistry and Chemical Technology, University of Ljubljana, Ve\v{c}na pot
113, 1000 Ljubljana, Slovenia}

\section{Introduction}

\ Since nearly two centuries ago van der Waals presented his equation of
state with the effective volume reduced by a finite size of molecules, hard
spheres were ubiquitously utilized to model gases, liquids and crystalls 
\cite{Feynman,Barrat,Mulero}, glasses \cite{Barrat,Parisi2010, Cond Matter
Review} and colloids \cite{RMP2024}, and many other condensed matter systems 
\cite{Fun}. The 3-dimensional (3D) hard sphere model has become the
statistical mechanical paradigm of a hard core repulsion and excluded
volume. Its 2D version, the hard disk (HD) model is a condensed matter
counterpart of the 2D Ising model in the magnetism. Both models are seen as
keys to understand and solve the 3-dimensional models, but in contrast to
the 2D Ising model, analytical results for the HD model are scarce. In this
situation, main information on a HD system is obtained by numerical methods 
\cite{Alder,squares,Kolafa,Krauth1,Krauth2,Krauth3}, which, in response to
the lack of theory, need to be continuously advanced \cite{JCP2021}. The
general difficulty in addressing hard core systems is that their interaction
has no small parameter and that finding the main thermodynamic potential,
entropy, is a novel purely geometrical prolem. The key idea is that the
entropy can be expressed in terms of the so-called free volume, which, in
the case of HD system, is area available for the disk center in a given
state of all other disks \cite%
{Barrat,Mulero,Hirschfelder,Newton,Rice,Hoover72,Hoover79,Speedy77,Speedy81,1995,1997,1998,1999,2006,2014,2015,2016,2023}%
. However, this idea has not been converted to the thermodynamics of even a
HD system. The reason is that the complex shape and connectedness of the 2D
free space make its precise measurement difficult, and the proposed
geometrical constructions \cite{1995,1997,1998,1999,2006,2014,2015,2016,2023}
have failed to exactly determine free volume \cite{Billiard}. These studies
were focused on its extensive fraction growing with the total volume, the
so-called cavity. \ Based on Widom's insertion method \cite{Widom}, cavity
in an $N$ disk system was associated with finding space for another disk in
the equilibrium $N-1$ disk system. But even for modest densities, cavities
become so rare that finding them was a task futile \cite{1998,Mulero}. At
the same time, Hoover, Ashurst, and Groover \cite{Hoover72} pointed out that
the free volume also has an intensive fraction, a private cell, related to a
single disk, which is nonzero even if cavity vanishes. This, however, has
not resolved the insertion problem as the extensive and intensive free
volume fractions were addressed only on separate grounds.

Thus, as free volume has not been established, its connection to the
partition function (PF) and entropy, however natural it be, remains a
hypotheses. The general difficulty is that the HD interaction has zero
energy, is "kinetic" rather than "potential" \cite{Feynman}, and the ten
known terms of the virial expansion behave incorrectly for the crystal
densities \cite{virial10}. As a result, the problem was addressed by very
different approaches. Considering the HD distribution related to the $%
\lambda $-point in liquid helium, Feynman discussed the problem
qualitatively by sketching mobile circles \cite{Feynman}. The essential
connection with HDs in the modern glass theory motivated Parisi and Zamponi
to develop the mean field theory for hard spheres in high unphysical
dimensions, from which one can apparently descent to the 2D and 3D \cite%
{Parisi2006,Parisi2010}. In this situation, since the Alder and Wainwright
discovery of the phase transition \cite{Alder} in 1962, the full equation of
state of the HD model has been obtained only from numerical experiments \cite%
{squares,Kolafa} (see review \cite{JCP2021} and references therein).

In this paper, using the language of multiple disks' intersections, we
obtain the following results. The exact formula for the free volume is
derived in terms of the intersection areas (IAs) of 2, 3, 4, and 5 disks of
twice the hard-core size; these IAs and thus the free volume can be computed
analytically as functions of disks' coordinates in the equilibrium system.
The free volume is shown to determine the PF integral and the entropy which
is the main thermodynamic quantity of the HD model. The $N$ disk PF $Z$ is a
product of the free volumes for all disk numbers from $1$ up to $N$ and has
two exact asymptotic expressions, $Z_{G}$ and $Z_{L}$. The $Z_{G}$ is
associated with the "gas" approximation (GA) when free volume is extensive,
the $Z_{L}$ with the\ "liquid" approximation (LA) when free volume is small
and intensive, and in between, $Z_{G}$ crossovers into $Z_{L}$. Our
numerical results have been obtained from disks coordinates generated by a
Monte Carlo simulation. Making use of their values, we analytically computed
the disks' IAs, PF, and finally the entropy. By the entropy differentiation,
the pressure has been computed for packing fractions $\eta $ up to the close
packing $\eta _{cp}=0.907$ and validated by a comparison with the numerical
dependence $P_{exp}(\eta )$ known from computer experiments \cite%
{Kolafa,squares}. For $\eta \lesssim 0.53$, $P_{exp}$ \cite{Kolafa} is
fitted by the $Z_{G}$, while for $0.72\lesssim \eta <\eta _{cp}$, $P_{exp}$ 
\cite{squares} is fitted by the $Z_{L}$. In the intermediate range, $%
0.53\lesssim \eta <0.69$, the system is a mixture of HDs at actual $\eta $
and defects in the form of HDs caged at $\eta _{d}=0.68$. These are
precursors of the nascent hexagonal order and, in line with the
Kosterlitz-Thouless scenario \cite{KT}, a source of entropy. Here the $%
P_{exp}(\eta )$ is recovered by the $Z_{L}$ for the mixture. A scalar order
parameter of the hexagonal order is identified with the IA of five disks. 
\begin{figure}[tbp]
\includegraphics[width=0.4\textwidth]{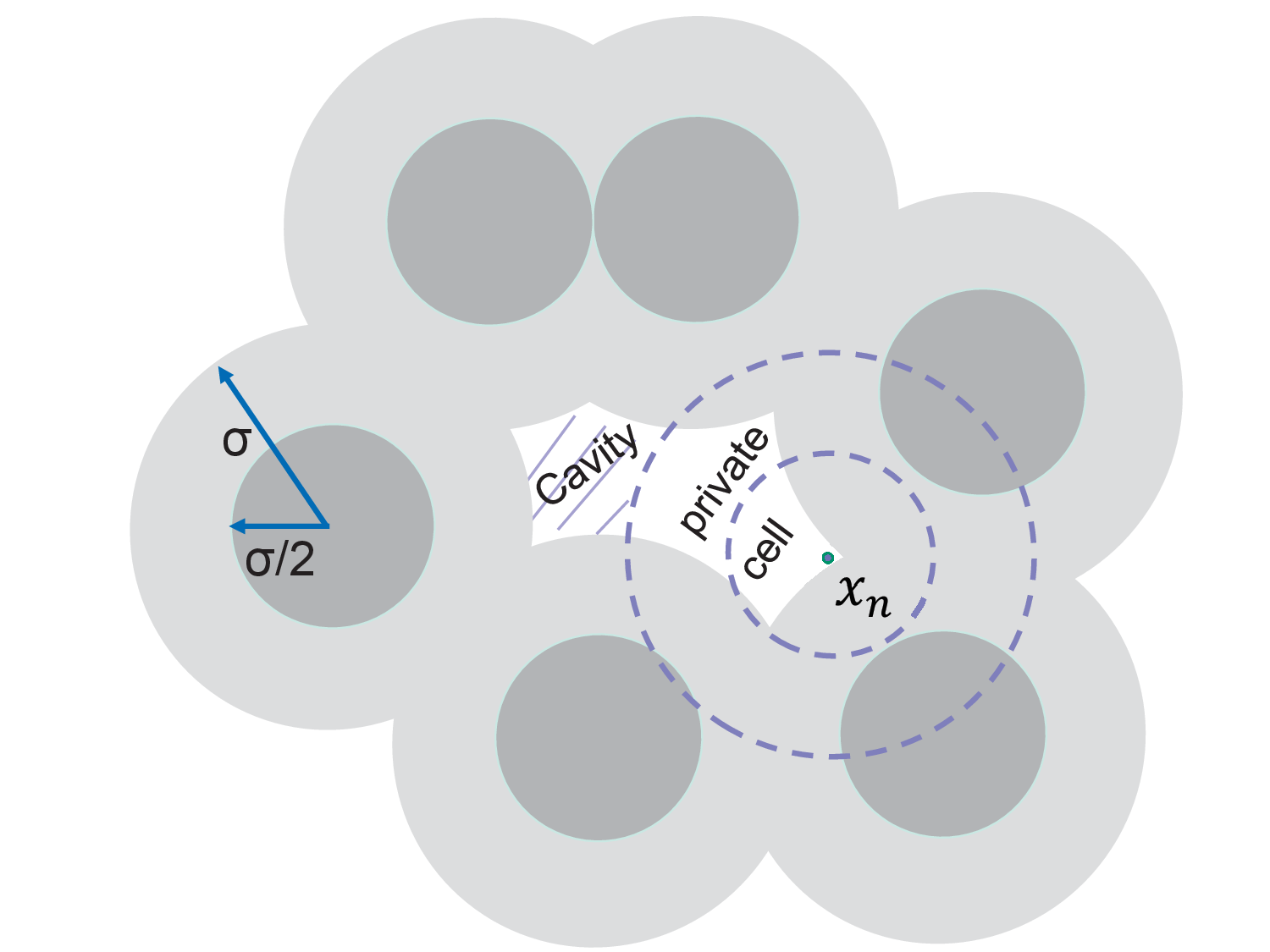} 
\caption{{}Fragment of a system of $N$ HDs. The $N-1$ HDs are dark circles,
and the connected $\protect\sigma $ circles are light circles. The $n$ th
disk and circle are shown by dashes. The inner white area is the free volume
of $n$ th disk, the hatched fraction is its cavity, and the clear fraction
is its private cell.}
\label{Fig1}
\end{figure}

\section{Free volume and multiple disk intersections}

For a given configuration $\{x\}_{N}=\{x_{1},...,x_{N}\}$ of $N$\emph{\ }HDs
with the center coordinates $x_{i}$, contained in the area $V,$ we define
several quantities, Fig.1. A HD of radius $\sigma /2,$ the core, is
supplemented with a concentric circle of radius $\sigma $ which is called
here $\sigma $-circle; $\sigma _{i}$ is a $\sigma $-circle centered at $%
x_{i} $. No disk center can enter $\sigma $-circles of other disks so that
their union forms an excluded volume. While hard cores cannot intersect, the 
$\sigma $-circles can. The free volume \emph{\ }$V_{N,n}$ of a disk\emph{\ }$%
n $ is the area accessible for its center $x_{n}$, i.e., the total $V$ minus
the area $\mu $ of the union of other $N-1$ $\sigma $ circles: 
\begin{equation}
V_{N,n}=V-\mu \left( \cup _{i\neq n}^{N}\sigma _{i}\right) .  \label{FV}
\end{equation}%
However, to avoid the insertion problem encountered in previous works, we
deal only with our system of $N$ HDs and also use a different, but
mathematically identical way to define $V_{N,n}.$ Fig. 1 shows that\ $%
V_{N,n} $ is indeed the sum of two quantities, the extensive cavity $C_{N}$
(hatched) and intensive private cell (clear) $c_{N,n}:$ 
\begin{equation}
V_{N,n}=C_{N}+c_{N,n}.  \label{Vn}
\end{equation}%
The cavity $C_{N}$\emph{\ }is the total $V$ minus area of the union of all $%
N $ disks: 
\begin{equation}
C_{N}=V-\mu \left( \cup _{i=1}^{N}\sigma _{i}\right) \,.  \label{CN}
\end{equation}%
The $C_{N}$, which can be described as the area where an "external" disk can
be inserted, can comprise more than one disconnected pieces and thus is
extensive. The private cell $c_{N,n}$ of a disk $n$ is the free volume of
this disk $V_{N,n}$ without the cavity, shown as the hatched area in Fig.1.
It is clear that this area is the $\sigma $-circle without areas of its
intersections with all its neighbors, i.e., 
\begin{equation}
c_{N,n}=\pi \sigma ^{2}-\mu \left[ \sigma _{n}\cap (\cup _{i\neq
n}^{N}\sigma _{i})\right] ,  \label{cNn}
\end{equation}%
which is obviously intensive (not connected to the total $V$). The division
of the free volume of disk $n$ in $C_{N}$ and $c_{N,n}$ depends on $x_{n}$,
but the total free volume $V_{N,n}$ does not. While $C_{N}$ (\ref{CN}) is
symmetric in all $\sigma _{i}$, the private cell $c_{N,n}$ (\ref{cNn}) of
disk $n$ essentially depends on\ its coordinate $x_{n}$ which will be
stressesed by an index $n.$ If $C_{N}=0$ and no additional HD can be
inserted, then the cell $c_{N,n}$ is the total free volume of the disk $n$.
Of course, $C_{N}$, $c_{N,n},$ and $V_{N,n}$ depend on the configuration $%
\{x\}_{N},$ which is not indicated for simplicity.

The above quantities can be expressed in terms of the IA of up to five $%
\sigma $-circles; intersections of more $\sigma $-circles without
intersection of their hard cores is impossible, Supplemental Material. The
IA of two $\sigma $-circles, centered at $x_{n}$ and $x_{i}$, is the area $%
\mu _{ni}$ common to both $\sigma $'s; similarly, $\mu _{nij},\mu _{nijk},$
and $\mu _{nijkl}$ are the IA of three, four, and five $\sigma $-circles
respectively. Now we introduce all $k$ disk IA $\mu _{k,n}$ involving disk $%
n:$ $\mu _{2,n}=\sum_{i}^{^{\prime }}\mu _{ni},$ $\mu
_{3,n}=\sum_{i>j}^{^{\prime }}\mu _{nij},$ $\mu
_{4,n}=\sum_{i>j>k}^{^{\prime }}\mu _{nijk},$ $\mu
_{5,n}=\sum_{i>j>k>l}^{^{\prime }}\mu _{nijkl},$ where prime indicates that
all summation indices are different from $n$. In terms of $\mu _{n,k}$, the
second term in (\ref{cNn}) is given by the known formula of the set theory,
and one has 
\begin{equation}
c_{N,n}=\pi \sigma ^{2}-(\mu _{2,n}-\mu _{3,n}+\mu _{4,n}-\mu _{5,n}).
\label{c1}
\end{equation}%
Using the same formula, the second term in (\ref{CN}) can be computed as the
sum of all IAs of each $\sigma _{n}$ with all other $\sigma $'s, regarding
the fact that, in this sum, IA of each two, three, four, and five $\sigma $%
's appears respectively two, three, four, and five times. Then 
\begin{equation}
C_{N}=V-\left[ N\pi \sigma ^{2}-\sum_{n=1}^{N}\left( \frac{\mu _{2,n}}{2}-%
\frac{\mu _{3,n}}{3}+\frac{\mu _{4,n}}{4}-\frac{\mu _{5,n}}{5}\right) \right]
.  \label{C1}
\end{equation}%
The above formula prompts that $C_{N}\{x\}_{N}$ can be conveniently
expressed in terms of the sums over disks' coordinates in the state $%
\{x\}_{N}$. We introduce the mean values $\mu _{k}$ which are $\mu _{k,n}$
of disk $n$ averaged over all $N$ disks in a given single state $\{x\}_{N}:$%
\begin{equation}
\mu _{k}=\frac{1}{N}\sum_{n=1}^{N}\mu _{k,n},k=2,3,4,5,  \label{muk}
\end{equation}%
In terms of these mean IAs, the cavity, mean private cell $\left\langle
c_{N}\right\rangle $, and mean free volume $\left\langle V_{N}\right\rangle $%
, which are the averages over all disks in a state, have the form%
\begin{eqnarray}
C_{N} &=&V-N\left[ \pi \sigma ^{2}-\left( \frac{\mu _{2}}{2}-\frac{\mu _{3}}{%
3}+\frac{\mu _{4}}{4}-\frac{\mu _{5}}{5}\right) \right] ,  \label{C2} \\
\left\langle c_{N}\right\rangle &=&\pi \sigma ^{2}-(\mu _{2}-\mu _{3}+\mu
_{4}-\mu _{5}),  \label{cN} \\
\left\langle V_{N}\right\rangle &=&C_{N}+\left\langle c_{N}\right\rangle .
\label{VN}
\end{eqnarray}

The $\left\langle c_{N}\right\rangle $ and $\left\langle V_{N}\right\rangle $
will naturally appear later while $C_{N}$ is a mean quantity by its
definition (\ref{C1}) and does not need the bracket. In Supplemental
Material, we present formulae which allow one to compute the IAs
analytically as functions of disks coordinates $\{x\}_{N}$ \cite{SM}. There
we also prove an important property that both $C_{N}$ and $c_{N,n}$
identically vanish exactly at the close packing (packing fraction $\eta =\pi
\sigma ^{2}N/4V)$. Below we show that the PF of the HD model is determined
by the free volume, which confirms that it is indeed the key quantity.

\section{The partition function and entropy}

In the thermodynamic limit and even in a system of large number of
particles, PF is practically an infinite-dimensional integral which cannot
be computed directly even numerically. The problem of statistical mechanics
is to solve this integral, i.e., to somehow reduce the infinite number of
actions (calculations) to a finite number of some other actions \cite{Baxter}%
. This is trivial in the case of ideal gas as then computing the PF reduces
to that of a certain finite-dimensional integral. The HD model is not of
this kind and we have only the second choice. In this section, we show that,
via the free volume, the PF of the HD model can be expressed in terms of
just four functions of density, i.e., $\mu _{k}$ (\ref{muk}), which can be
computed by the above analytical formulae using disks' coordinates. Thus,
establishing the PF reduces to computing of the four functions of $\eta $ 
\cite{Analytical theory}.

The interaction between HDs $i$ and $j$ of the radius $\sigma /2$ is $%
U_{ij}=\infty $ for $x_{j}$ within $\sigma _{i}$ and $U_{ij}=0$ for $x_{j}$
outside $\sigma _{i}$. The configurational PF of $N$ HDs in the volume $V,$ $%
Z\propto \int_{V}dx_{1}...dx_{N}\exp \left( -\sum_{i<j}^{N}U_{ij}\right) ,$
up to a factor determined below can be presented in the form%
\begin{widetext}
\begin{eqnarray}
Z&=&\int_{V}dx_{1}\int_{V}dx_{2}e^{-U_{21}}\int_{V}dx_{3}e^{-U_{31}-U_{32}}...\int_{V}dx_{k}e^{-
\sum_{j=1}^{k-1}U_{kj}}...\int_{V}dx_{N}e^{-\sum_{j=1}^{N-1}U_{Nj}}
\label{ZNN} 
 \\
&=&\int_{V}dx_{1}\int_{V_{2}(x_{1})_{N}}dx_{2}\int_{V_{3}(x_{1},x_{2})_{N}}dx_{3}...\int_{V_{k}\{x_{k-1} 
\}_{N}}dx_{k}...\int_{V_{N}\{x_{N-1}\}_{N}}dx_{N}.  
\notag
\end{eqnarray}
\end{widetext}The $k$ th integration domain $V_{k}\{x_{k-1}\}_{N}$ is shown
in Appendix A, eq.(A2), to be the free volume of $k$ th disk in the $N$ disk
system, which is not within the union of chosen $k-1$ $\sigma $ disks $%
\sigma _{i},i=1,2,...,k-1:$ $V_{k}\{x_{k-1}\}_{N}=$ $V-\mu \left( \cup
_{i=1}^{k-1}\sigma _{i}\right) _{N}$ , where $\{x_{k-1}\}_{N}$ is a
subconfiguration of $k-1$ disks of the full $N$ disk configuration $%
\{x\}_{N}.$ For instance, by definition (\ref{FV}), $V_{N}\{x_{N-1}\}_{N}$
is the standard free volume in the $N$ th disk in the $N$ disk system. As $%
V_{k}\{x_{k-1}\}_{N}$ does not explicitly depend on the $x_{k^{\prime }}$
with $k^{\prime }>k$, it is instructive to identify allowed $\{x_{k-1}\}_{N}$
as a configuration, which leaves space in $V$ for the rest $N-k+1$ disks
irrespective of their specific coordinates. Thus, as allowed $%
\{x_{k-1}\}_{N} $ are subject to this accommodation condition, the $%
V_{N}\{x_{N-1}\}_{N}$ is a conditional free volume, but for brievity we call
it just free volume. For a high density $N/V$, this accommodation condition
is a restriction on $\{x_{k-1}\}_{N}$, which is addressed below.

The central result, proven in Appendix A, is that the general PF $Z$ (\ref%
{ZNN}) factorizes into the product of the free volumes $V_{k}\{x_{k-1}\}_{N}$
averaged over all allowed $\{x_{k-1}\}_{N}$ $:$%
\begin{figure}[tbp]
\includegraphics[height=2 in]{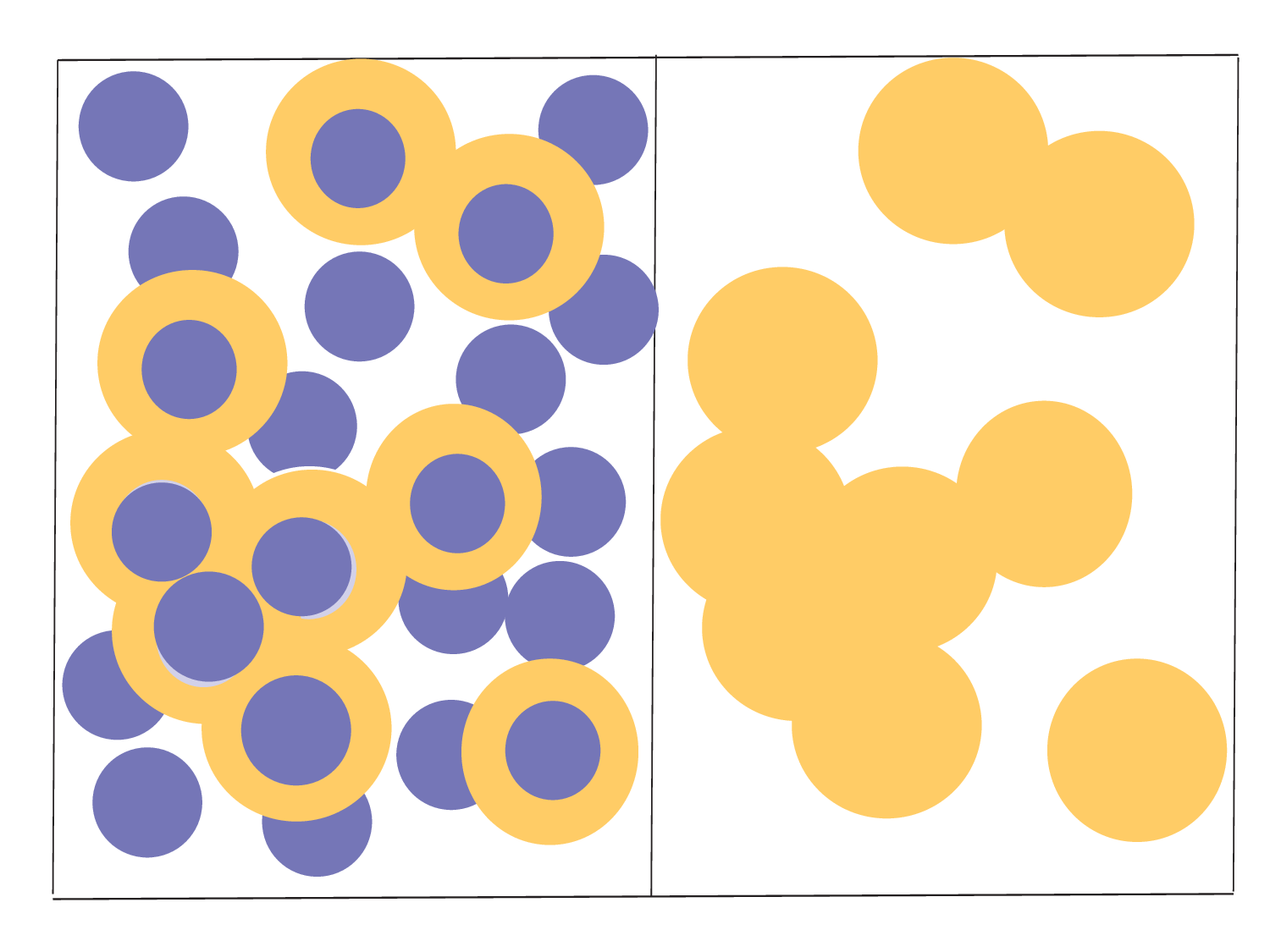} 
\caption{{}Left panel. A system of $N=23$ HDs with dark cores. The chosen $%
k-1=9$ HDs have light concentric $\protect\sigma $-circles. Right panel. The
light $\protect\sigma $ circles are those in the left panel and the empty
area is the free volume $V_{10}\{x_{9}\}_{23}$ for any dark disk chosen as $%
10$ th disk. As here $\protect\eta $ is below 0.5, the rest $N-k+1=14$ HDs
are accommodated for any positions of the chosen 9 disks, the cavity
dominates in the free volume and the GA is justified.}
\end{figure}
\begin{equation}
Z\propto \prod_{k=1}^{N}\left\langle V_{k}\right\rangle _{N},  \label{ZZ}
\end{equation}%
where the subscript $N$ indicates the above accommodation condition.
However, for a packing $\eta $ below $0.5$, the empty space in $V$ is in
average larger than its occupied complement, so that all $N$ HDs can be
accommodated for any positions of $k-1$ disks, Fig.2. In this "gas"
approximation (GA), the integration domain is that of just $k$ disks in $V$.
In the GA, the free volume is unconditional and dominated by the extensive
cavities, disks can exchange their positions producing identical states, and
the division of the PF by $N!$ is in order. Regarding this property in (\ref%
{ZNN}), the PF $Z_{G}$ in the GA takes the form%
\begin{equation}
Z_{G}=\frac{1}{N!}\prod_{k=1}^{N}\left\langle C_{k}\right\rangle ,
\label{ZG}
\end{equation}%
where index $N$ is of no need. There is a close similarity of the above idea
behind the computing $\left\langle C_{k}\right\rangle $ to the one used to
compute permutation of $N$ objects: object $k$ sees objects $%
k-1,k-2,...,2,1, $ but does not reflect on the positions of the rest $N-k$
objects. As a result, the formula (\ref{ZG}) can be seen as a continuous
version of factorial (this relation is addressed also in Appendix A).
Obviously, for $k\ll N,$ $\left\langle V_{k}\right\rangle =$ $V-k\pi \sigma
^{2}+O(\pi \sigma ^{2}/V),$ but for large $k$, the $\left\langle
V_{k}\right\rangle $ is determined by the IAs $\mu _{2,3,4,5}$ from formula (%
\ref{C2}) with $N$ replaced by $k$. \ Introducing the densities $\rho
_{k}^{\prime }=k/V$ and $\rho =N/V,$ the configurational gas entropy $%
S_{G}=\ln Z_{G}$ can be converted to the integral :%
\begin{equation}
S_{G}=V\int_{0}^{\rho }d\rho ^{\prime }\ln \left\langle C(\rho ^{\prime
})\right\rangle -N\ln N/e.  \label{SG}
\end{equation}

The standard idea of $Z$ implies the integrations over the maximum areas,
e.g., the $x_{1}$ integration is over the total $V$, the $x_{2}$ integration
is over $V$ without $\sigma _{1}$, and so on. This is an idealization as
actual areas, explored by particles over the relaxation time $\tau _{rel}$,
are much smaller. This is true for any many-body system so that the
configurational entropy is determined at best up to a constant. However, for
a sufficiently low density, local cavities are connected and during $t_{rel}$
disks swap their positions. Then the integration can be formally extended
over total $V$ while the factor $1/N!$ compensates the double counting.
However, \ as $\eta $ increases, local cavities get disconnected and disks
cannot travel beyond their cages. Then over $t_{rel}$, which is now very
short, disks can explore only their private cells $c$ and all $V_{k}$ tend
to their mean common value $\left\langle c_{N}\right\rangle .$ This common
value is exactly $\left\langle c_{N}\right\rangle $ introduced in (\ref{cN}%
), Appendix A.

In this approximation which we call liquid (LA), the free volumes are
strictly conditional and shrink into private cells, division by $N!$ has no
sense as disks do not exchange their positions, and the PF $Z_{G}$ and
entropy $S_{G}$ crossover into 
\begin{eqnarray}
Z_{L} &=&\left\langle c_{N}\right\rangle ^{N},  \label{ZL} \\
S_{L} &=&N\ln \left\langle c_{N}\right\rangle ,  \label{SL}
\end{eqnarray}%
Above we estimated that the GA is expected to work for $\eta $ below $\sim
0.5$. Similarly, the LA range is for $\eta $ above $\sim 0.5$. In this
range, the occupied fraction of $V$ is larger than the empty complement,
accommodation of $N-k$ disks essentially depends on the $k$ disks'
distribution, which imposes a restriction on $\{x_{k}\}_{N}$. Of course, in
between, a transition from the GA to LA is expected, but this general
problem is yet to be addressed \cite{Hexatic}. Our calculations of the free
volume show that the cell contribution is negligible in the GA for $\eta
\lesssim 0.53$ and cavities are negligible in the LA for $\eta \gtrsim 0.58$%
, so that the crossover range is very short. In the next section, we compute
the pressure in the GA and LA and compare it with the $P_{\exp }.$

\begin{figure}[tbp]
\includegraphics[width=0.5\textwidth]{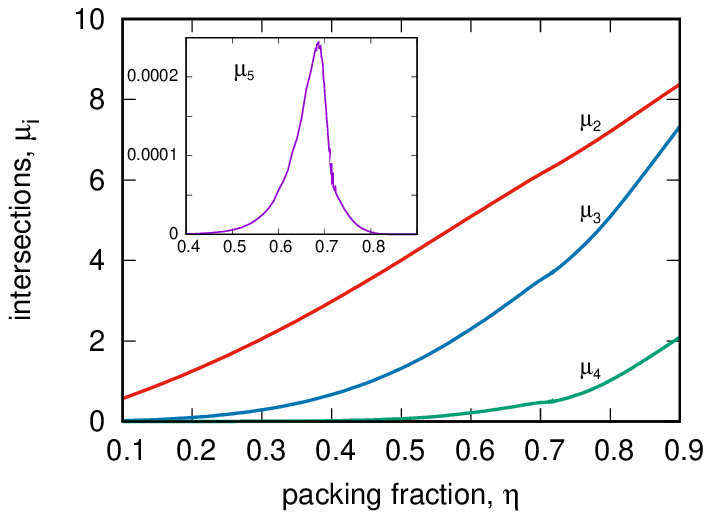}
\caption{{} The mean IAs $\protect\mu _{2}$, $\protect\mu _{3}$, and $%
\protect\mu _{4}$ of two, three, and four $\protect\sigma $-circles as
functions of $\protect\eta $. Inset: the mean IA $\protect\mu _{5}$ of five $%
\protect\sigma $ circles. The diameter $\protect\sigma =1$.}
\label{Fig3}
\end{figure}

\begin{figure}[tbp]
\includegraphics[width=0.5\textwidth]{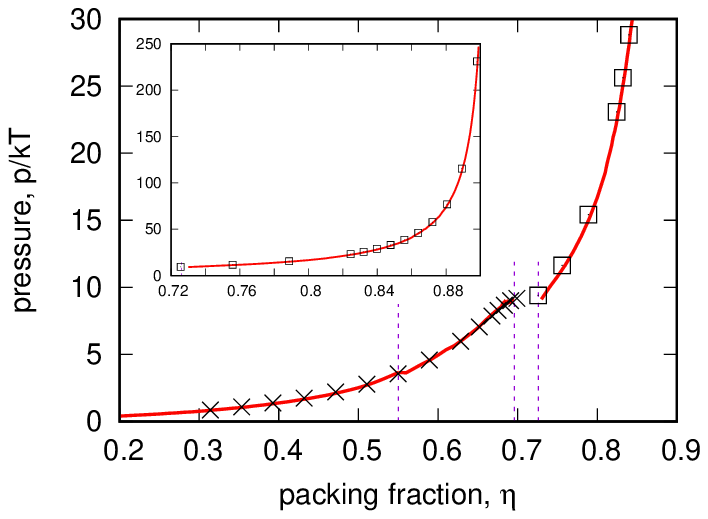}
\caption{{} Theoretical pressure $P(\protect\eta )$ (solid) superimposed on
the simulation result $P_{exp}$ (crosses \protect\cite{Kolafa} and squares 
\protect\cite{squares} ). The vertical dash lines delimit, from left to
right, the gas region, crossover (with a small mismatch a bit rightward of $%
\protect\eta =0.55$) and mix-liquid region, the phase coexistence (empty),
and liquid region. }
\label{Fig4}
\end{figure}

\begin{figure}[tbp]
\includegraphics[angle=-90,width=0.5\textwidth]{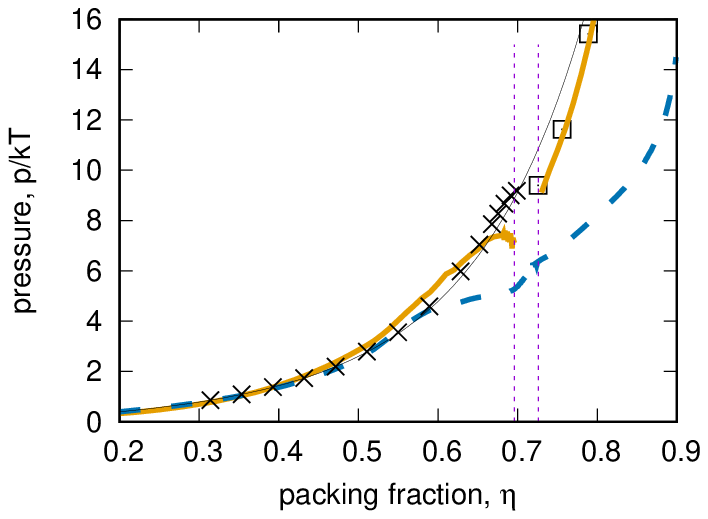}
\caption{{} The pressures $P_{exp}$ (symbols), $P_{L}$ in the LA (solid), $%
P_{G}$ in the GA (dash), and pressure resulting from the ten virial terms
(thin solid)\protect\cite{virial10}. }
\label{Fig5}
\end{figure}

\section{Recovering the "experimental" pressure $P_{exp}$: gas, liquid, and
in between}

\textit{The gas and liquid regions.} As indicated in Appendix A and as the
formulae for the free volume show, in the thermodynamic limit or just for a
sufficiently large $N,$ the free volume and the main observables of the HD
model can be obtained from disk coordinates in a single equilibrium
configuration. In the present work, such microscopic input has been
generated by Monte Carlo simulations of a system containing $900$ HDs in the
range $0.1<\eta \leq 0.9$. Next for a given packing fraction $\eta $ we analytically computed
the IAs to get the configuration mean IA $\mu _{k}$, $k=2,3,4,5$ (%
\ref{muk}) (the diameter $\sigma =1$) which are presented in Fig.3 and will be addressed later. Then
these functions were used to compute the cavity contribution $C_{N}$ (\ref%
{C2}), the mean private cell $\left\langle c_{N}\right\rangle $ (\ref{cN}),
and the entropies $S_{G}$ (\ref{SG}) and $S_{L}$ (\ref{SL}). The pressure
was obtained by their differentiation with respect to the volume $V=N/(4\eta
/\pi ),P_{G,L}=\partial S_{G,L}/\partial V$ ($k_{B}T$ is set one), and
compared with the well-established simulation equation of state $P_{\exp }$
reported in Refs. \cite{Kolafa,squares}. We found that the $S_{G}$ and $%
S_{L} $ allow one to recover the $P_{\exp }$ in the range of packing
fractions up to $\eta _{cp}$ with a gas-liquid crossover region $%
0.53\lesssim \eta \lesssim 0.58,$ Fig.4. We do not address only the
well-known range $0.69\lesssim \eta \lesssim 0.73$ of two phase coexistence 
\cite{Krauth1,Krauth2} as the phase separation, necessary for computing $\mu
_{n}$ in each of them, requires a highly advanced simulation with $N\sim
10^{6}$ \cite{Krauth1,Krauth2,JCP2021}. Note that the pressure $P_{L}$ is a
smooth function of $\eta $ everywhere, but in the phase coexistence range
random oscillations suddenly appear and disappear exactly at its end points,
and the reason is that here the IAs are taken for randomly distributed
phases.

We found four distinctive regions of $P(\eta )$, which we conditionally call
gas, crossover, mix-liquid, and liquid region, Fig.4 \cite{Hexatic}. In the
lower density gas region, $\eta \lesssim 0.53$, the $P_{\exp }$ is well
fitted by $P_{G}$ of the GA. The $P_{L}$ gives a very good fit to $P_{\exp }$
in the liquid region, $\eta \gtrsim 0.726,$ and diverges exactly at $\eta
_{cp}$. Fig.5 shows that, in line with our qualitative $\eta $ range
division between GA and LA, above $\eta $ $\sim 0.53$ the $P_{G}$ starts to
deviate from $P_{\exp }$ and then at $\eta _{cp}$ diverges as $\ln
\left\langle V_{N}\right\rangle $ whereas $P_{\exp }$ diverges as $%
1/\left\langle V_{N}\right\rangle $. There are two regions between $\eta
\sim 0.5$ and $\eta \simeq 0.69.$ In between $\eta \sim 0.53$ and $\eta \sim
0.58$, the GA crossovers into the LA as expected. In the mix-liquid region, $%
0.58\lesssim \eta \lesssim 0.69,$ the fit to $P_{\exp }$ can be obtained in
the LA, but here HDs are mixed with defects.

\textit{In between gas and liquid: the mix-liquid region. }In the mix-liquid
region, the $P_{L}$ does not fit to $P_{\exp }$, Fig.5. For $\eta >0.55,$ $%
P_{L}$ first goes above $P_{\exp }$ indicating that as $\eta $ increases,
the actual entropy $S_{L.m}$ decreases slower than $S_{L}$ and hence is
higher than $S_{L}$. The additional entropy gain can be due to a mixing with
foreign particles or defects. Analyzing the HD configurations, Huerta et al 
\cite{Huerta} revealed that, \ above $\eta \sim 0.5$, defects in the form of
a HD caged by their neighbors at the density $\eta _{d}=0.68$ appear in the
HD configurations, their fraction $\zeta $ grows with $\eta $ and tends to $%
1 $ at the phase coexistence. Such caged disks are precursors of the nascent
ordering at $\eta _{d}=0.68.$ As such a dense disks' clump in the form of
defect cage appears at a different point, the configuratiion of a system at
the actual lower density changes and the entropy gains. If $\zeta N=N_{d}$
is the number of defects, the PF of the HD-defect mixture is 
\begin{equation}
Z_{mix}=(N-N_{d})Z_{L}(\eta )\cdot N_{d}Z_{L}(\eta _{d})\cdot \lbrack
N!/N_{d}!(N-N_{d})!],  \label{Zmix}
\end{equation}

and the entropy $S_{L,mix}$ is%
\begin{eqnarray}
S_{L,mix} &=&N[(1-\zeta )\ln \left\langle c_{N}(\eta )\right\rangle +\zeta
\ln \left\langle c_{N}(\eta _{d})\right\rangle  \label{Smix} \\
&&-(1-\zeta )\ln (1-\zeta )-\zeta \ln \zeta ].  \notag
\end{eqnarray}

\begin{figure}[th]
\includegraphics[angle=-90,width=0.5\textwidth]{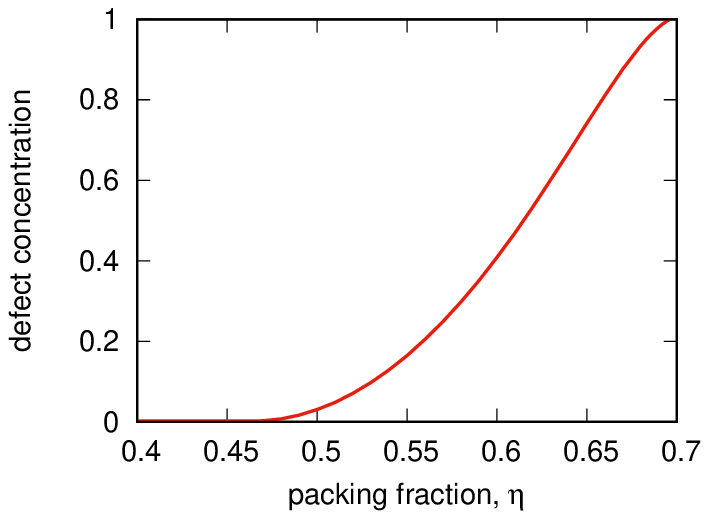}
\caption{{}Phenomenological density of caged defects $\protect\zeta(\protect%
\eta)$ as a function of \ packing fraction $\protect\eta .$}
\label{Fig2}
\end{figure}
Here $\zeta (\eta )$ is the fraction of caged disks at $\eta $ that must
vanish below $\eta \sim 0.5$ and is nearly one at $\eta =0.69.$ The $P_{\exp
}$ is well fitted by $\partial S_{L,m}/\partial V$ for $\zeta (\eta )$ shown
in Fig.6. Thus, in the intermediate range, HDs at the actual $\eta $ are
mixed with precursors of the nascent hexagonal order approaching at $\eta
=0.68$ \cite{Huerta}. As in the Kosterlitz-Thouless scenario \cite{KT},
generating defects the system increases its entropy. The entropy $S(\eta )$,
smoothly interpolating those in the different regions, is presented in
Fig.7. \ 
\begin{figure}[tbp]
\includegraphics[angle=-90,width=0.5\textwidth]{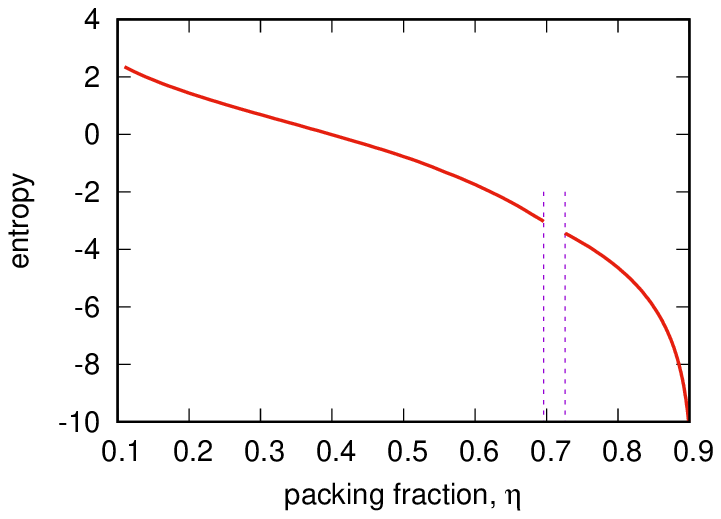}
\caption{{} Theoretical entropy continuously interpolating $S_{G}$, $%
S_{L,mix}$, and $S_{L}$.}
\end{figure}
\ \qquad \qquad \qquad

\section{Intersecsion of five $\protect\sigma $-circles and the hexagonal
order}

Now we address behavior of the mean IAs $\mu _{2},\mu _{3},\mu _{4},$ and $%
\mu _{5},$ Fig.3. The IA of two, three, and four $\sigma $-disks are
monotonically growing functions of $\eta ,$ which is natural as the disk
separations are getting shorter and shorter. However, the $\mu _{5}(\eta )$
is very different: it increases, attains maximum at the beginning of the
phase coexistence $\eta _{5,\max }=0.687$, and then monotonically decreases
to zero at $\eta _{cp}.$ This can be understood in connection to the
hexagonal order. First, it is not difficult to see that at the perfect order 
$\mu _{5}$ vanishes: even in the close packed hexagonal lattice, the IA of
five $\sigma $-disks is that of a single point and can be nonzero only if
the order is not perfect, see Supplemental Material. Second, its
compression-induced increase terminates just at the onset of phase
coexistence where the ordering is supposed to start \cite%
{Krauth1,Krauth2,Krauth3,Wang}. We see that although $\mu _{5}$ is much
smaller than the other $\mu _{k}$, it is a natural indicator of the local HD
ordering. In other words, the function 
\begin{equation}
h(\eta )=1-\mu (\eta )/\mu _{5,\max },  \label{h}
\end{equation}%
\begin{figure}[th]
\includegraphics[angle=-90,width=0.5\textwidth]{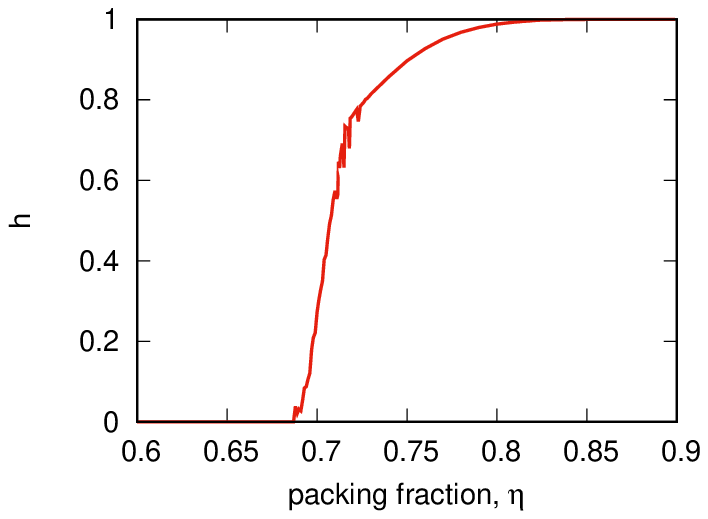}
\caption{{}Local scalar order parameter $h$ (\protect\ref{h}) of the
hexagonal order as a function of packing fraction $\protect\eta $. The
irregularity in $h$ in the range $0.69\lesssim \protect\eta \lesssim 0.73$
is due to the two phase coexistence as explained in the text.}
\label{Fig8}
\end{figure}
Fig.8, where $\mu _{5,\max }=\mu _{5}(0.687)\simeq 2.4507\cdot 10^{-4},$ can
be used to describe the scalar component of the hexagonal order along with
its orientational component \cite{Hexatic}. This is similar to the case of a
nematic liquid crystal which has both orientational and scalar order
parameters. An even closer analogy can be drawn with a smectic liquid
crystall which also has a 2D quasi-long range global orientational order 
\cite{Landau} with the local scalar order parameter \cite{Mukherjee}. Thus, $%
h$ shows that, as $\eta $ increases from $0.687$, the hexagonal order
monotonically tends to the perfect lattice at the close packing $0.907$,
Supplemental Material.

\section{Discussion and conclusion}

The set intersections, unions, and their measures are an essential tool of
the ergodic theory which enables\ one to deal with\ a phase space of
arbitrary complex shape \cite{KSF}. The free volume is an example of a set
with such a complex shape. The significance of the five disk intersection $%
\mu _{5}$ along with the successful description of the equation of state in
terms of the free volume show how natural the language of $\sigma $-circle
intersections is for the statistical geometry of a HD system. The free
volume, which incorporates the extensive cavity and intensive private cell
on common ground in the actual system of $N$ HDs, was obtained in terms of
the multi-circle IAs $\mu _{2}(\eta ),\mu _{3}(\eta ),\mu _{4}(\eta ),$ and $%
\mu _{5}(\eta ),$ eq.(\ref{muk}), which can be computed analytically, given
the disks' coordinates. Free volume is the key quantity as it determines the
PF and entropy and thus the thermodynamics of the HD model: solving the
infinite-dimensional PF for each density reduces to knowing the disks
coordinates or the four functions $\mu _{k}$. Computation-wise, it is
crucial that the accuracy is restricted only by that of disks' coordinates
as calculation of the IAs is analytical. These results were obtained for the
first time and validated by comparison with the known simulation equation of
state $P_{\exp }(\eta )$. In particular, we found a Kosterlitz-Thouless-type
effect of entropy gain due to the defect creation in the range $0.58\lesssim
\eta \lesssim 0.69$ preceding the phase coexistence. Our theory raises the
following important problems. First, to find the defect density in the
mix-liquid region which in this paper was addressed phenomenologically.
Second, to smoothly describe the GA to LA crossover. Third problem concerns
the four functions $\mu _{k}$ of $\{x_{N}\}.$ As these functions have
already been computed once and for all, Fig.3, and as they determine the PF
and entropy, they themselves can be used to determine the thermodynamics of
the HD model \cite{Analytical theory}. Derivation of these functions is a
well-posed mathematical problem which in principle can be addressed without
resorting to disk coordinates. Moreover, these four functions contain much
less information than the sets of disk coordinates for different $\eta $,
hence they represent the minimum information required for establishing the
thermodynamics of the HD model and thus appear to be the most elementary
bits of the statistical geometry for a given $\eta $. At last, the free
volume is a general geometrical quantity that universally appears in the
problems of any hard core particles, and our approach in terms of disk
intersections paves a general way to address them. In particular, our
formulae for the PF and free volume can be directly adopted for the hard
sphere model \cite{HS}. Of course, in systems with hard core shapes more
complex than a disk or sphere, finding the intersection volumes is possible
only numerically and can be computationally demanding. However, the
intersection approach introduced in this paper has the following fundamental
advantage: rather than computing and summing an infinite series, the free
volume, however complex it be, and thus the PF can be obtained exactly by
computing a finite number of intersections.

\section{Acknowledgements}

VMP is grateful to the Center for Theoretical Physics PAS for hospitality.
This research is part of the project No. 2022/45/P/ST3/04237 co-funded by
the National Science Centre and the European Union Framework Programme for
Research and Innovation Horizon 2020 under the Marie Sk\l odowska-Curie
grant agreement No. 945339. T.B. was supported by the National Research
Foundation of Ukraine Project No. 2023.05/0019. A.T. acknowledges financial
support through the MSCA4Ukraine project, funded by the European Union
(grant agreement ID: 1039539).%


\appendix{}

\section{Appendix A: Partition function in terms of the free volume.}

Here we first show that the integration domains in (\ref{ZNN}) are the free
volumes and then prove the central result (\ref{ZZ}): the PF factorises into
their product. Consider $I_{k}=\int_{V}dx_{k}\exp \left(
-\sum_{i=1}^{k-1}U_{ik}\right) ,$ the $k$ th integral in the general
expression for the PF $Z$ (\ref{ZNN}). The potential $U_{ik}$ is such that
the center $x_{k}$ of a circle $\sigma _{k}$ cannot enter any $\sigma _{i}$
with $i<k$. The circles $\sigma _{i}$,\ introduced in the main text, can be
considered as sets of points $x_{k}$ $:$ $\sigma _{i}=\{x_{k}:\left\vert
x_{i}-x_{k}\right\vert \leq \sigma \},i<k.$ By definition, indicator $\tau
_{i}$ of a set of points $x_{k}\in \sigma _{i}$ is $\tau _{i}=1$ for $%
x_{k}\in \sigma _{i}$ and $\tau _{i}=0$ otherwise. The product of $n$
indicators of $n$ different $\sigma _{i}$ is indicator of the intersection
set $\bigcap_{i=1}^{n}\sigma _{i}$, shared by all of them: $\tau _{1}\tau
_{2}...\tau _{n}=\tau \left( \bigcap_{i=1}^{n}\sigma _{i}\right) $. Noting
that $\exp (-U_{ik})=1-\tau _{i},$ the exponential in $I_{k}$ can be
expressed as a product of such terms :%
\begin{equation}
\exp \left( -\frac{1}{2}\sum_{i=1}^{k-1}U_{ik}\right)
=\prod\limits_{i=1}^{k-1}(1-\tau _{i})  \tag{A1}
\end{equation}%
\begin{eqnarray*}
&=&1-\sum_{i=1}^{k-1}\tau _{i}+\sum_{i>j}^{k-1}\tau _{i}\tau
_{j}-\sum_{i>j>l}^{k-1}\tau _{i}\tau _{j}\tau _{l}+... \\
&&+(-1)^{k-1}\sum_{i_{1}>i_{2}>...>i_{k-1}}^{k-1}\tau _{i_{1}}...\tau
_{i_{k-1}}.
\end{eqnarray*}%
As more than five $\sigma $ circles cannot intersect without intersection of
their cores, all products of more than five $\tau $'s do not contribute to
the above sum. Now we integrate both sides of (A1) over $V$ regarding the
definitions of $\mu _{k,i}$ in the text and eqs.(\ref{c1}) and (\ref{CN}) to
get 
\begin{equation}
I_{k}=V_{k}\{x_{k-1}\}_{N}.  \tag{A2}
\end{equation}%
This is the free volume of a $k$ th disk in a maze made by a configuration $%
\{x_{k-1}\}_{N}$ of $k-1$ disks taken from the total $N$ HDs. This allowed $%
\{x_{k-1}\}_{N}$ is subject to the following accommodation condition: it
must leave space in $V$ for the rest $N-k+1$ disks, \textit{no matter} what
their configuration is, Fig.2. Let us show that the general PF $Z$ (\ref{ZNN}%
) is the product of the mean free volumes. To start, we set $k=N$ and
introduce the integral operator 
\begin{equation}
Z_{N-1,N}=\int_{V}^{^{\prime }}dx_{1}...dx_{N-1}\exp \left(
-\sum_{i>j}^{N-1}U_{iN}\right) ,  \tag{A3}
\end{equation}%
where the prime indicates that the integral is over all \textit{allowed}
configurations $\{x_{N-1}\}_{N}$ of $N-1$ disks (i.e., they leave a space
for disk $N$). It is important that $Z_{N-1,N}$, which is apparently
reminiscent of Widom's PF $Z_{N-1}$ of $N-1$ HDs in $V$ \cite{Widom}, is not 
$Z_{N-1}$ as the $\sigma _{N}$ is accommodated in $V$ at some $x_{N}$ and
acts on the other disks. Indeed, it is excluded from the integration domain
in (A3) \ as $U_{iN}$ is the hard core repulsion of disks $i$ by disk $N$,
Fig.7 \cite{insertion problem}. Regarding (A2), the integral $Z$ (\ref{ZNN})
can be presented as the action of $Z_{N-1,N}$ on $V_{N}:$ 
\begin{equation}
Z\propto Z_{N-1,N}\{x_{N-1}\}_{N}(V_{N}\{x_{N-1}\}_{N}).  \tag{A4}
\end{equation}

\begin{figure}[tbp]
\includegraphics[height=2 in]{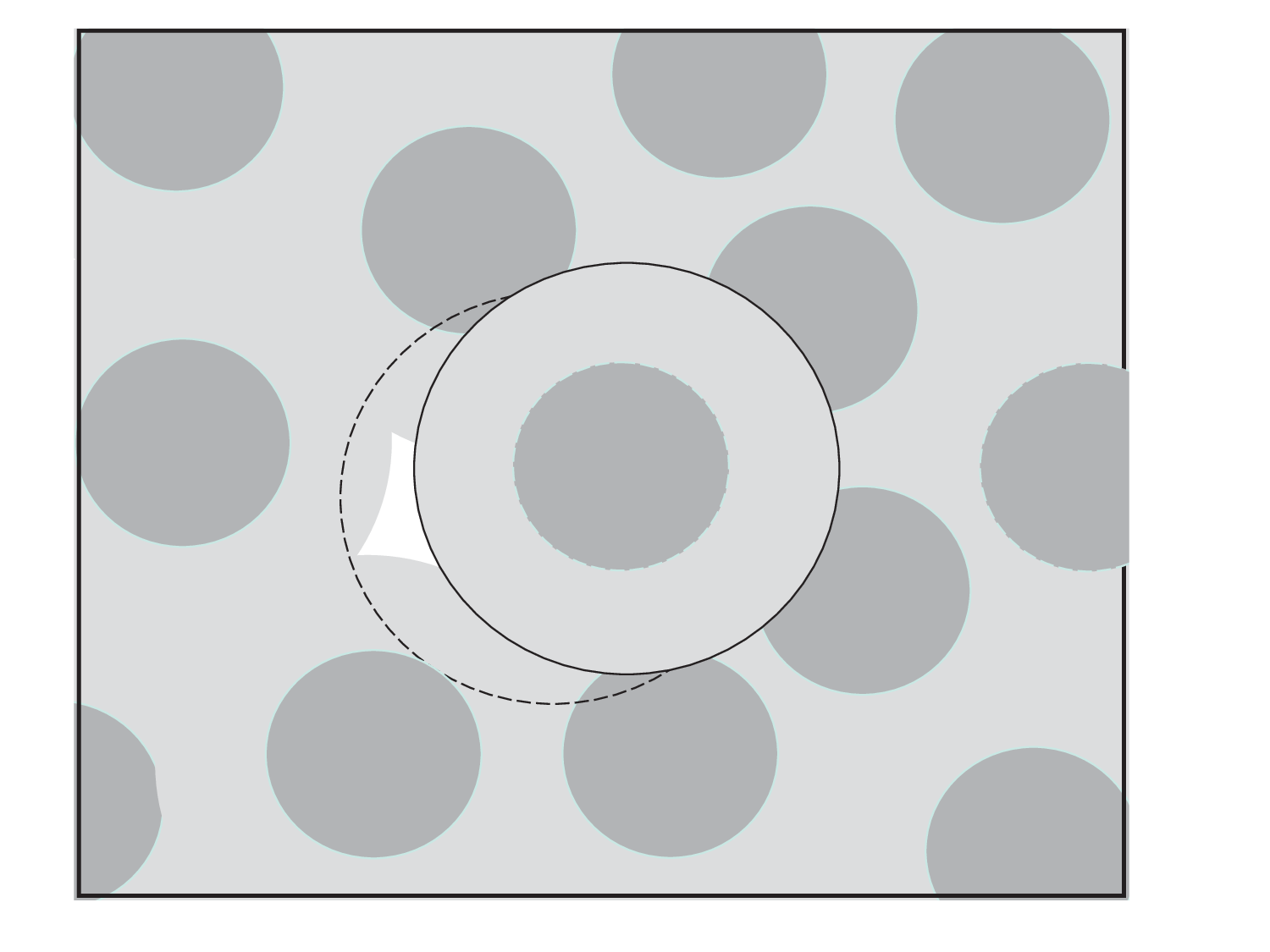}
\caption{{}A configuration of $N-1$ HDs in an $N$ disk system. It is allowed
as its $\protect\sigma $ circles leave a cell (white area) for another, $N$
th disk. However, if considered as equilibrium configuration of an $N-1$
disk system as in Widom's approach, then it is more homogeneous with the
central circle moved to the dashed position. This is a not allowed
configuration as the $N$ th disk has no place in it.}
\label{Fig8}
\end{figure}

A HD system has the statistical property of mixing \cite{Sinai}. The mixing
is much stronger than ergodicity: while ergodicity provides a convergence to
the time average, in a mixing system an extensive observable $O$ itself
converges to its limiting value $\left\langle O\right\rangle $ in the
thermodynamic equilibrium \cite{KSF}. In the equilibrium, the system moves
from one equilibrium $\{x\}$ to another equilibrium $\{x^{\prime }\}$, but
in a large system almost all $\{x\}$ (by the phase space measure) give this
value, $O\{x\}=O\{x^{\prime }\}=\left\langle O\right\rangle $. At the same
time, for each equilibrium $\{x\}_{N}$, an observable $\left\langle
O\right\rangle _{N}$ $=$ $\left\langle \left\langle O\right\rangle
\right\rangle _{N}$, where $\left\langle \left\langle O\right\rangle
\right\rangle _{N}$ can be considered as the expectation value in an
infinite ensemble. This is shown below.

Consider a single equilibrium $\{x\}_{N}$ in the thermodynamic limit $%
N,V\rightarrow \infty ,$ $N/V=\rho .$ Divide the infinite volume $V$ into,
e.g., $\sqrt{N}$ equal $\Delta V_{i}$ of size $V/\sqrt{N}$ with the number
of disks $\sqrt{N}+o(\sqrt{N})\approxeq \sqrt{N}$ and density $\rho =N/V$.
In all $\Delta V_{i}$ actual arrangements of disks are different and
therefore they represent an infinite number of different realizations of
configurations of same number of disks and density in similar infinite
systems. If $O_{\sqrt{N},i}$ is the value of $O$ in the volume in $\Delta
V_{i},$ then it follows that $\left\langle \left\langle O\right\rangle
\right\rangle _{N}=\sqrt{N}\left( \sum_{i=1}^{\sqrt{N}}\left\langle
O\right\rangle _{\sqrt{N},i}/\sqrt{N}\right) =\sqrt{N}\left\langle O_{\sqrt{N%
}}(\rho )\right\rangle =\left\langle O\right\rangle _{N}$ and thus $%
\left\langle O_{N}(\rho )\right\rangle =\left\langle \left\langle O_{N}(\rho
)\right\rangle \right\rangle .$

Now we proceed with the PF $Z,$ eq.(A4). The free volume $%
V_{N}\{x_{N-1}\}_{N}$ is the sum of cavity $C\{x_{N-1}\}_{N}$ (cavity of a
disk in a configuration of $N-1$ disks subject to the condition that they
leave a place for $N$ th disk) and the private cell $c_{N}\{x_{N-1}\}_{N}$
(private cell of $N$ th disk in a configuration of $N-1$ disk). Since, as we
stressed in the main text, the last has a peculiar dependence on the
coordinate of disk $N,$ it must be considered separately from $C_{N}.$
Consider first the contribution of $C$ to $Z$ (A4). By virtue of mixing, the
value of $C\{x_{N-1}\}_{N}$ is the same for each equilibrium $%
\{x_{N-1}\}_{N} $ so that 
\begin{equation}
Z_{N-1,N}\{x_{N-1}\}_{N}(C\{x_{N-1}\}_{N})=C_{N}\times Z_{N-1,N}.  \tag{A5}
\end{equation}%
Now consider $c_{N}\{x_{N-1}\}_{N}.$ Its contribution to $V_{N}$ depends on
the coordinate $x_{N}$ and specific $\{x_{N-1}\},$ but the total sum over
all $\{x_{N-1}\}$ of course does not. To find this constant value $%
\left\langle c_{N}\right\rangle $ we notice that the result will be same if $%
x_{N}$ is swaped with any $x_{i},i=1,2,...,N-1.$ Then 
\begin{eqnarray*}
&&Z_{N-1,N}(c_{N})+Z_{N-1,N}(c_{N-1})+...+Z_{N-1,N}(c_{1}) \\
&=&N\times Z_{N-1,N}(\left\langle c\right\rangle ),
\end{eqnarray*}%
whence $\left\langle c_{N}\right\rangle =(1/N)\sum c_{n,N},$ the expected
result which is formula (\ref{cN}). Combining this with (A5), the right hand
side of (A4) becomes $\left\langle V_{N}\right\rangle _{N}\times Z_{N-1,N}$
where $\left\langle V_{N}\right\rangle _{N}=C_{N}+\left\langle
c_{N}\right\rangle .$ Continuing along this line for $k=N-1,N-2,...$ one
obtains $Z$ (\ref{ZNN}) in the form of a product of the free volumes,
determined by the $k-1$ disk subconfigurations of the actual $N$ HD
configurations for all $k=1,2,...,N$ :%
\begin{equation}
Z\propto \prod_{k=1}^{N}\left\langle V_{k}\{x_{k-1}\}_{N}\right\rangle
=\prod_{k=1}^{N}\left\langle V_{k}\right\rangle _{N},  \tag{A6}
\end{equation}%
which is eq.(\ref{ZZ}). The proportionality factor depends on the
approximation, which in turn depends on the packing. As explained in the
main text, for a low "gas" $\eta $, the factor is $1/N!,$ $\left\langle
V_{k}\right\rangle _{N}$ can be replaced by $\left\langle C_{k}\right\rangle
,$ and $Z=Z_{G}$ (\ref{ZG}). For a "liquid" density $\eta >0.55,$ all $%
\left\langle V_{k}\right\rangle _{N}$ reduce to private cells, the factor is
unity, and $Z=Z_{L}$ (\ref{ZL}).

In the main text we briefly introduced the idea that PF is the integral over
the states which are accessible, explored, passed by the system over the
relaxation time $t_{rel}$. This is natural as over another $t_{rel}$ the
system reproduces its equilibrium state and thus doubles the number of
states explored over the first $t_{rel}$. We have shown that, through the
free volume, this idea can provide a continuous transition from the GA to
LA, thus enabling a single theory of the HD model in the entire range of
densities. Here we address this important idea in more detail.

It is important to realize that the standard PF in the form (\ref{ZNN}) or
(A6), if treated formally as the integral over the whole volume $V$, does
not tend to the PF (\ref{ZL}) of the LA and thus cannot give the correct
behavior for $\eta $ above $0.72$ (actually, above $0.53$). The key to
seeing this is in observing that, if treated formally, expression (A6) is a
continuous generalization of a factorial and intimately related to the
permutations and following lattice model: $N$ particles on $J$ lattice
sites, no more than one on each site of size $V/J$, where $V$ is the total
volume. The exact PF of this model is $(V/J)^{N}J!/(J-N)!N!$ and the
pressure $P_{J,N}=-(N/V)ln[(V/N)(1-N/J)]$. At the close packing, $%
N\rightarrow J$, this pressure diverges as $ln($free volume$)$, the same way
as the pressure of the GA and the virial pressure \cite{virial10}, Fig.5,
whereas the correct pressure, both $P_{exp}$ and $P_{L}$, diverges as $1/($%
free volume$)$. The lattice model shows that a logarithmic divergence is the
general corollary of the integration over the progressively decreasing
areas, starting from the full volume in $Z$ (A6) and $Z_{G}$ (\ref{ZG}), and
over all $J$ sites in the lattice model. Such integration presupposes that,
in the actual system state, configurations of disks and those, obtained by
swaps of their coordinates, have equal probabilities. This is pertinent to
the "gas" densities, when the cavity consists of large connected pieces
which are accessible for disks and where they can swap their positions
during relaxation time $t_{rel}$. Then the integration is formally extended
to the entire space allowing all permutations, while the double counting is
compensated by the factor $1/N!$. But at the "liquid" densities, disks
vibrate in the cages made by their neighbors, and position swapping requires
a breaking from one cage to another cage. This event is highly improbable
during the time $t_{rel}$ which is now very short. Thus, swapping is a very
seldom event, a fluctuation whose probability is negligible in a large
system. Hence disks do not permute and there is no double counting to take
care of.

In the above partition of $V,$ there always is one subvolume $\Delta V$
which contains disk $N$ and its private cell $c_{N,N}$ vanishing only at the
close packing. And this is inspite of a possibility that, for high $\eta ,$
there might be no cavities in all $\Delta V_{i}$ (or a negligible number of
fluctuations in form of a cavity in some $\Delta V_{i}$ ) and $\left\langle\left\langle
C_{N}(\rho )\right\rangle\right\rangle =0$. We see that, if $\eta <\eta _{cp}$, a finite
free volume is guarantied for every HD. Thus, addressing both extensive and
intensive free volume fractions on the same footing and considering only the
original system of $N$ HDs, one resolves the following paradox of the disk
insertion method: a disk, that belongs to the system, cannot be inserted
into this very system and thus has no place in it \cite{insertion problem}.

\pagebreak


\onecolumngrid
\section{\textbf{Supplemental Material}}
\section{Vanishing of the cavity and intensive cell in a densely packed
triangular lattice}

Here the procedure of counting and computing areas of all possible multiple
intersections of the $\sigma $ circles for a single disk and computing the
cavity and cell volume is demonstrated for the close packed triangular
lattice which has the single configuration and does not need averaging. A
fragment of this lattice is shown in Fig.10a. The disks' cores, which are in
contact with each other, are filled and have radius $\sigma /2;$ the
attached concentric $\sigma $ circles of radius $\sigma $ are shown by
dashes (shown only for five disks). The central disk $0$ is shown along with
its six next neighbors, $1,2,3,4,5,6,$ and with its six next next neighbors $%
7,8,9,10,11,12$; the distance of these next next neighbors to $0$ is less
than $\sigma $ so that their $\sigma $ circles can overlap with the central $%
\sigma $ circle. The upper fragment with the five shown $\sigma $ circles is
sufficient for the establishing of all neighbor $\sigma $ circles
overlapping with the central circle because it is one of the three similar
fragments. No other disks in the lattice have their $\sigma $ circles
overlapping with $0$ circle.

\begin{figure}[th]
\includegraphics[height=10cm]{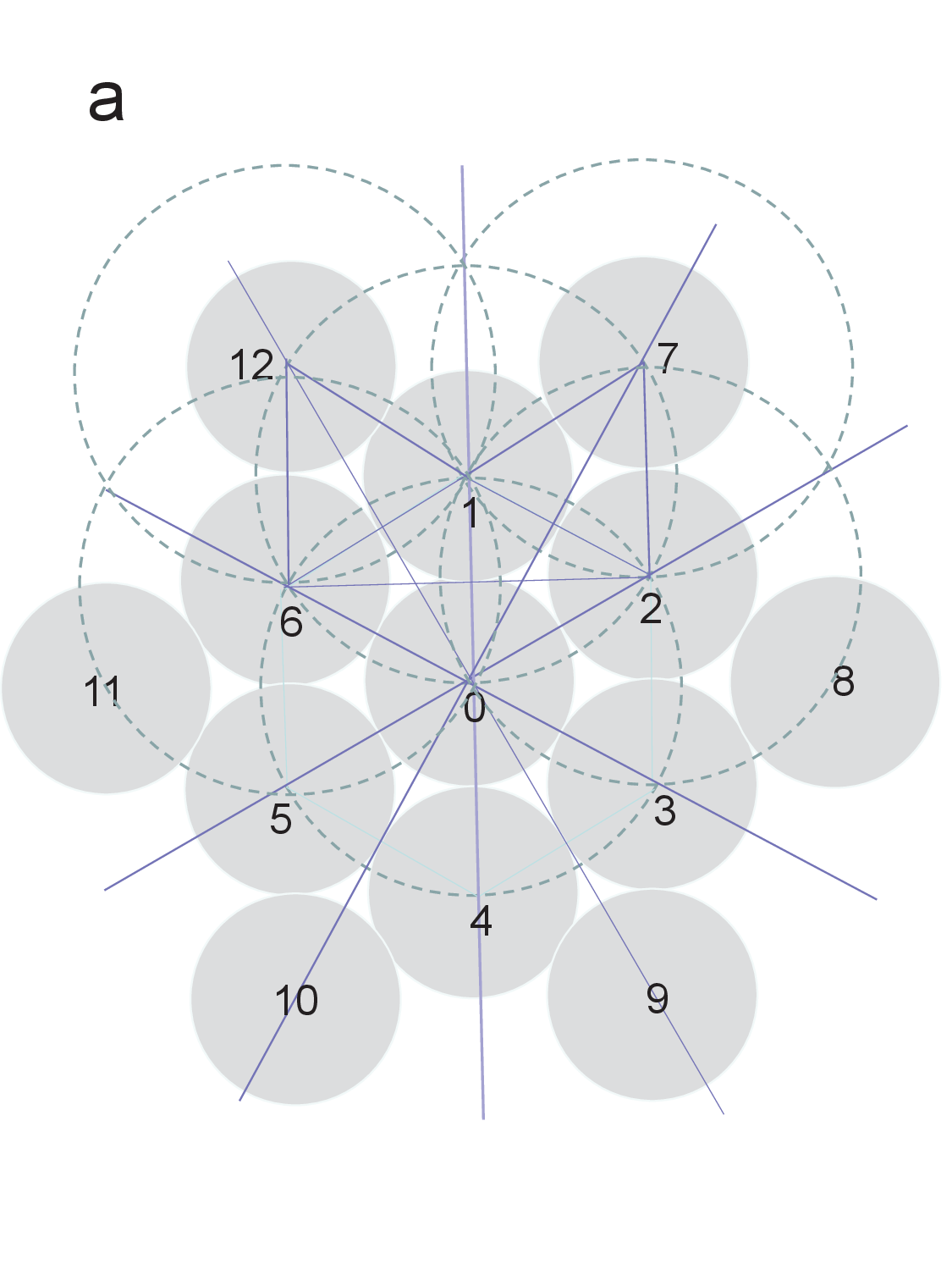} %
\includegraphics[height=9cm]{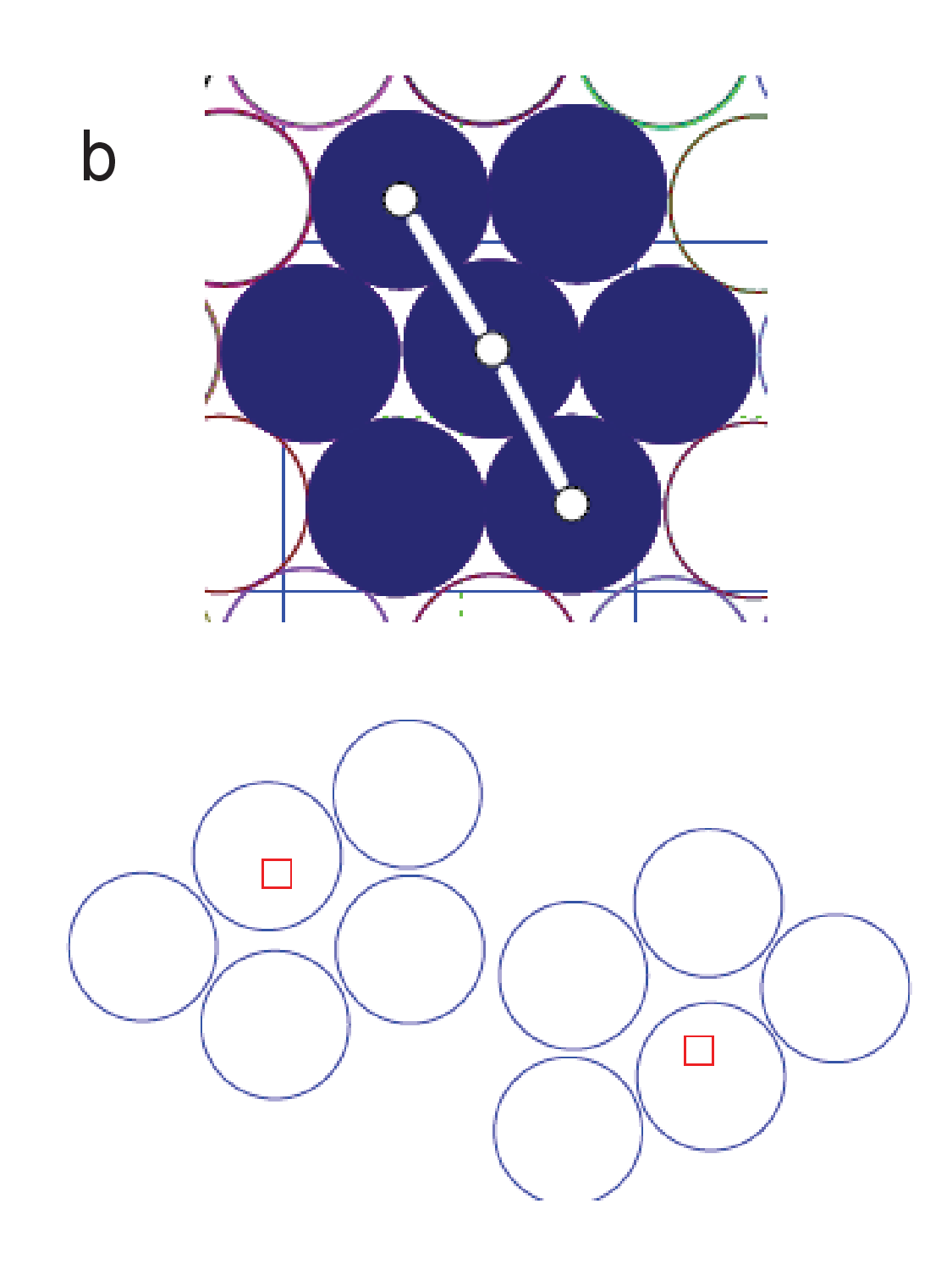} 
\caption{{}a. A fragment of triangular densely packed HD lattice. The disks'
cores are filled circles of the radius $\protect\sigma /2.$ Five $\protect%
\sigma $ circles centered at $0,$ 2, 7$,$ 12, and 6 are indicated by dashes.
The central disk $0$ has next neighbors centered at 1,2,3,4,5, and 6, and
next next neighbors centered at 7,8,9,10,11, and 12. \ $\protect\sigma $
circles of all disks in the lattice, which are not shown, do not overlap
with the $\protect\sigma $ circle centered at $0$. The intersection area of
five disks, e.g., 02165, 02834, is a point. b. Five disk intersection area
can be finite only if the hexagonal order is violated. In the upper panel,
the white line is not straight as it would be in a perfect triangular
lattice, and the intersection area of the five disks on the left of the line
is nonzero. In the lower panel, the intersection areas of the two sets of
five disks are shown by the squares.}
\label{Fig1}
\end{figure}

$S_{2}.$ First we notice that, in the perfect lattice, five $\sigma $
circles intersect in a single point: Fig.1b illustrates that five $\sigma $
circles can have a nonzero intersection area only if the perfect hexagonal
order is violated. Next, for the circles indicated by dashes, we find and
list different $\sigma $ pairs, $\sigma $ triples, $\sigma $ quadruples,
which include $0$ circle; then we compute their surface areas and finally
count the total numbers of these different terms. There are six pairs of
circles similar to $01$ and six pairs similar to $07,$ which gives for the
total contribution of pairs $S_{2}=6S_{01}+6S_{07}.$ The $S_{01}$ is the
area $0216$ bounded by circle 0 from above and circle 1 from below, its area
is $S_{01}=2(\pi /3-\sqrt{3}/4)\sigma ^{2};$ the area $S_{07}$ is the lobe
with the vortices $1$ and 2, $S_{07}=2(\pi /6-\sqrt{3}/4)\sigma ^{2}.$

$S_{3}.$ Similarly, $S_{3}=6S_{102}+12S_{017}+6S_{602}=6S_{102}+18S_{017}$
as $S_{602}=S_{017}.$ The area $S_{102}$ is that of the curvelinear triangle 
$102$, $S_{102}=[\sqrt{3}/4+3(\pi /6-\sqrt{3}/4)]\sigma ^{2};$ another
tripple area $S_{017}$ is that of the lobe with the vertices $1$ and 2, so
that $S_{017}=S_{07}=2(\pi /6-\sqrt{3}/4)\sigma ^{2}$.

$S_{4}.$ $S_{4}=6S_{0216}+6S_{0172}=12S_{0216}.$ Finally, the area $S_{0216}$
is the lobe with vertices 0 and 1 which is equal to $S_{07},$ $%
S_{0216}=2(\pi /6-\sqrt{3}/4)\sigma ^{2}.$

Now we are ready to compute the close packing values of the intensive cell
volume $c_{cp}=\pi \sigma ^{2}-\mu _{2}+\mu _{3}-\mu _{4}+\mu _{5}$ and the
cavity $C_{cp}=V-N\left[ \pi \sigma ^{2}-\mu _{2}/2+\mu _{3}/3-\mu
_{4}/4+\mu _{5}/5\right] $ using the formulae (6) and (5) of the main text.
Substituting the above values of $S_{n},$ one finds: 
\begin{equation}
c_{cp}/\sigma ^{2}=\pi -\left[ 6(\pi /6-\sqrt{3}/4)+2\pi -3\sqrt{3}/2\right]
=\pi -\pi =0.  \label{O1}
\end{equation}%
The cavity is extensive and to deal with the size independent quantities, we
devide $C_{cp}$ by $N\sigma ^{2}.$ The expression $V/N\sigma ^{2}=\pi
/(4\eta _{cp})$ where $\eta _{cp}=N\pi \sigma ^{2}/4V$ is the packing
fraction at close packing, $\eta _{cp}=2\pi /\sqrt{3}$. Substituting the
above values of $S_{n}$ in $C_{cp}$ one gets:%
\begin{equation}
C_{cp}/N\sigma ^{2}=\pi /4\eta _{cp}-\pi +\pi -\sqrt{3}/2=0.  \label{O2}
\end{equation}%
It is essential that not only the total free volume, but both intensive and
extensive free volumes vanish separately which shows their functional
independence. It is also important to realize that, as evident from Fig.2, a
small increase in the disks' separation will results in nonzero intensive $c$
whereas the extensive $C$ will remain zero. Actually, a simple calculation
using the formulae, given in this Supplemental Material, shows that the
cavity in a perfect triangular lattice vanishes already at $\eta =0.393.$

\section{Excluded volume for a system of HDs of radius $\protect\sigma /2:$
Intersections of 2,3,4, and 5 disks of radius $\protect\sigma $.}

The general consideration of disk intersections was addressed by Kratky \cite%
{Kratky1,Kratky2} who obtained a number of elegant and useful results.
However, for real applications to computing intersection areas $\mu
_{3,4,5}$ from the HD coordinates, the general formulae obtained by Kratky
require a nontrivial and meticulous development related to their actual
highly conditional character. Here we present such formulae which we used in
our analytical computations of $\mu _{3,4,5}.$

Let $\sigma /2$ be the HD radius. Our effective $\sigma $-circles have
radius $\sigma $ and we need to find their intersections areas given their
cores of radius $\sigma /2$ do not overlap.

\subsection{Intersection of two HDs.}

The volume of intersection of two disks $m$ and $n$ is \cite{Kratky1}

\begin{equation}
I_{nm}=\left\{ 
\begin{array}{c}
2\sigma ^{2}\left\{ \cos ^{-1}(r_{nm}/2\sigma )-(r_{nm}/2\sigma )\left[
1-(r_{nm}/2\sigma )^{2}\right] ^{1/2}\right\} ,0\leq r_{nm}\leq 2\sigma );
\\ 
0,r_{nm}>2\sigma .%
\end{array}%
\right. .  \label{I2}
\end{equation}

The sum of all areas of pair intersections $\mu _{2,n}$ of the disk $n$ is%
\begin{equation}
\mu _{2,n}=\sum\limits_{m\neq n}I_{nm}.  \label{eta2}
\end{equation}

\subsection{Intersection I$_{mnl}$ of three HDs.}

Consider three disks $(m,n,l)$ from the $2\sigma $ vicinity of $o$
(including $o)$ such that $dist(m,n)\leq 2\sigma ,$ $dist(n,l)\leq 2\sigma
,dist(m,l)\leq 2\sigma ,$ for otherwise $I_{mnl}=0$ (only the intersections
with disk $n$\ are needed for its contribution to the excluded volume, but
tripples of disks without $n$ from the 2$\sigma $ vicinity of $n$ are needed
for the four and five disks' intersections including $n$). Consider a
triangle $\triangle (nml).$ The three disk intersection depends on whether
the circumradius $R_{mnl}$ exceeds or not exceed $\sigma .$

\textbf{1)}. $0\leq R_{mnl}<\sigma .$ Then the intersection is \cite%
{Kratky1,Kratky2}: 
\begin{equation}
I_{mnl}=\frac{1}{2}\left( I_{mn}+I_{ml}+I_{nl}-\pi \sigma ^{2}\right) +\frac{%
1}{4}\frac{r_{mn}\cdot r_{ml}\cdot r_{nl}}{R_{mnl}},  \label{I3}
\end{equation}%
where $R_{mnl}$ is the circumradius of $\triangle (nml):$ 
\begin{eqnarray}
R_{mnl} &=&\frac{r_{mn}\cdot r_{ml}\cdot r_{nl}}{\sqrt{%
p(p-2r_{mn})(p-2r_{ml})(p-2r_{nl})}},  \label{R} \\
p &=&r_{mn}+r_{mo}+r_{no}.  \notag
\end{eqnarray}

\textbf{2).} $R_{mnl}\geq \sigma .$ Then there are exactly two cases.

\ \ \ \ \ \ \ 2in) The in-case when the center of the circumcircle $C_{mnl}$
is \textit{inside} the triangle $mnl$ without its boundary (sides)$,$ 
\begin{equation}
I_{mnl}=0,C_{mnl}\in \triangle mnl.  \label{2in}
\end{equation}

\ \ \ \ \ \ 2out) $\ $The out-case\textbf{\ }when the center of the
circumcircle $C_{mnl}$ is outside the triangle $mnl$ (the "outside" includs
sides of $\triangle mnl,$ $\partial \triangle mnl).$ Then one first finds
the largest side of $\triangle mnl.$ Let $max(mn,ml,nl)=mn$ ($mn$ is both
length of side $mn$ and just the side $mn)$, then 
\begin{equation}
I_{mnl}=I_{mn},C_{mnl}\notin \triangle mnl,C_{mnl}\in \partial \triangle mnl.
\label{2out}
\end{equation}

This is \textit{a very important task} as all the intersections of 3, 4, and
5 $\sigma $ disks are determined by the presence of triangles with
circumradia larger than $\sigma .$ If you have the catalogue of all
triangles from the $2\sigma $ vicinity of disk $n$ $($with and without disk $%
n$ itself$)$ along with the positions (in or out) of the centers of their
circumcircles, then further calculations are considereably simplified and
accelerated.

The sum of all triple intersections including disk $n$ that contributes to $%
\mu _{3,n}$ is%
\begin{equation}
\mu _{3,n}=\sum\limits_{\forall \triangle (nml)}I_{mnl}.  \label{eta3}
\end{equation}

\subsection{Intersection of four HDs.}

Consider four disks, i.e., $n$ and a triple$~(k,m,l)$ assuming
that the distance between any pair made out of $n,m,l,k$ is $\leq 2\sigma .$
Then you consider a quadrangle consisting of four triangles, $knl$, $%
kml,nml,knm.$ Now we also need to know what disks stay at the diagonals of
the quadrangle $knml$. Suppose, that 
\begin{equation*}
max[nm,nl,nk,mk]=nm.
\end{equation*}%
Then $nm$ is the lagest diagonal and $kl$ is the second diagonal of $knlm$.
Again, the formulae are different if the circumradia of all the four
triangles are below $\sigma $ or if there is\textit{\ at list one triangle}
with circumradius above $\sigma .$

\textbf{1)} the circumradia of all the four triangles are below $\sigma $, $%
mn>kl$ \cite{Kratky1}: 
\begin{eqnarray}
I_{knlm} &=&I_{nkl}+I_{mkl}-I_{kl},  \label{I4} \\
\forall R^{\prime }s &<&\sigma .  \notag
\end{eqnarray}

\textbf{2}) \ One (or more) $R\geq\sigma .$ Practically, you check $R_{nkm}$
and $R_{nlm}$ related to the largest diagonal $mn.$ Then, if $R_{knm}\geq
\sigma $ and the center $C_{knm}\notin \triangle (nmk)$ ($C_{knm}\in
\triangle (knm)$)$,$ then 
\begin{eqnarray}
I_{knlm} &=&I_{mnl},R_{knm}\geq \sigma ,C_{knm}\notin \triangle (nmk);
\label{I44} \\
I_{knlm} &=&0,R_{knm}\geq \sigma ,C_{knm}\in \triangle (nmk).  \notag
\end{eqnarray}%
This $I_{mnl}$ shows why you need the catalogue of triangles including those
without $n.$ The sum of all four-disk intersections of the disk $n$ that
contributes to the excluded volume is%
\begin{equation}
\mu _{4,n}=\sum\limits_{\forall \square (knlm)}I_{knlm}.  \label{eta4}
\end{equation}%
Here $mn$ and $kl$ are necessarily the diagonals of quadrands $knlm$ (not $%
knml$ $!$)$.$

\subsection{Intersection I$_{knlm5}$ of five HDs.}

Now we take the quadrangles $I_{knlm}$ for which $I_{knlm}>0,$ and add to
each of them all possible fifth apexes $m^{\prime }=5$ to build all possible
pentagons that can contribute to the excluded volume. We remind the Reader
that in this case $kl$ and $nm$ are diagonals of the correspondent
quandrangle (this is essential!) and $mn>kl.$

\textbf{1}). If the circumradia of all the four triangles of $knlm$ are
below $\sigma ,$ the intersection of 5 disks is \cite{Kratky1}:%
\begin{equation}
I_{knlm5}=I_{mkl5}+I_{nkl5}-I_{kl5}.  \label{I5}
\end{equation}%
\ 

\textbf{2}). \ One (or more) triangles in the quadrangle $knlm$ have $%
R\geq\sigma .$ Let $R_{nlm}\geq\sigma $. If $C_{nlm}\notin \triangle (nlm)$ then $%
I_{onlmk}=I_{onm5}$ and the problem is reduced to the intersection of four
disks. Otherwise, i.e., if $C_{nlm}\in \triangle (nlm)$ then $I_{onlm5}=0:$ 
\begin{eqnarray}
I_{onlm} &=&I_{mnl},R_{nlm}\geq \sigma ,C_{nlm}\notin \triangle (nlm);
\label{I55} \\
I_{onlm} &=&0,R_{nlm}\geq \sigma ,C_{nlm}\in \triangle (nlm).  \notag
\end{eqnarray}

The general rule is that if, in the reduction of the intersection to a lower
number of intersections you find any zero intersection, then the total
intersection is zero.

The five-disk contribution due to the five disk intersections $I_{knml5}$ of
the disk that contribute to the excluded volume we need is$:$ 
\begin{equation}
\mu _{5,n}=\sum I_{knmlk}.  \label{eta5}
\end{equation}


\begin{thebibliography}{99}
\bibitem{Billiard} The complex shape of the free volume has a fully convex
boundary which allowed Sinai to prove that the HD system has the property of
K-mixing \cite{Sinai} and thus does relax to the (microcanonocal)
equilibrium state. In relation to the mixing, see Appendix A.
\bibitem{Analytical theory} This is a usual requisite of solutions of
many-body problems: solving an infinite-dimensional PF consists in
identification of a finite number of any quantities (e.g., eigenvalues of
certain transfer matrix usually obtained only numerically) that enable one
to get the PF \cite{Baxter}.
\bibitem{Hexatic} The numerical simulations of a 2D HD system \cite%
{Krauth1,Krauth2,Soft Matter,Krauth3} and experiments with 2D colloids \cite%
{Exper Colloids} have shown that its phase sequence reflects the
Kosterlitz-Thouless-Halperin-Nelson-Young scenario \cite{KTHNY}. In the
established HD phase diagram, a disordered liquid is below $\eta \simeq 0.7,$
above it up to the end of the phase coexistence at $\eta \simeq 0.72$, the
phase is hexatic with a quasi-long hexagonal orientational and no
translational order, and above $\eta \simeq 0.72$ it smoothly transitions to
a crystal with only a quasi-long range translational and apparently long
range orientational order. Nevertheless, in this paper it is more
instructive to be connected to the specific forms of the PF. The "gas"
approximation range, GA, where the PF has the standard form (\ref{ZG}), in
particular includes the actual HD gas densities. The range $\eta >0.58$
comprises both conventional liquid and solid, but we call it "liquid"
approximation, LA, since the PF here has the same form (\ref{ZL}). This form
of the PF was widely accepted in the phenomenological theory of liquids in
the thirties of the last century \cite{Hirschfelder,Newton,Rice} and later
in the cell models \cite{Hoover72,Hoover79,Huerta}. These theories remained
qualitative as finding the free volume was unsolved problem.
\bibitem{HS} The 3D case is more computationally demanding: the $\sigma $%
-circle is replaced by a sigma-ball of radius twice that of a sphere, the
number of $\sigma $-ball intersections without intersections of the hard
sphere cores is 11, whereas analytical approach has so far been developed
only for the volumes of up to four ball intersection \cite{4HS}. However, in
3D, these common volumes are expected to decrease with the number of balls
much faster than in the 2D.
\end{thebibliography}
\end{document}